# The Integration of High-k Dielectric on Two-Dimensional Crystals by Atomic Layer Deposition

Han Liu, Kun Xu, Xujie Zhang and Peide D. Ye [a]
School of Electrical and Computer Engineering and Birck Nanotechnology Center
West Lafayette, IN 47906, USA

## Abstract

We investigate the integration of $Al_2O_3$ high-k dielectric on two-dimensional (2D) crystals of boron nitride (BN) and molybdenum disulfide ($MoS_2$) by atomic layer deposition (ALD). We demonstrate the feasibility of direct ALD growth with trimethylaluminum(TMA) and water as precursors on both 2D crystals. Through theoretical and experimental studies, we found that the initial ALD cycles play the critical role, during which physical adsorption dominates precursor adsorption at the semiconductor surface. We model the initial ALD growth stages at the 2D surface by analyzing Lennard-Jones Potentials, which could guide future optimization of the ALD process on 2D crystals.

[a] Author to whom correspondence should be addressed; Electronic mail: yep@purdue.edu

The application of atomic layer deposition (ALD) techniques to metal gates and high-k dielectrics in the past decade has triumphantly extended Moore's Law for the continued scaling down of silicon based CMOS devices[1]. In addition, the integration of high-k materials on other semiconductors, such as Ge, GaAs, InGaAs, GaSb, etc., has also been comprehensively studied in the pursuit of alternative channel materials to replace silicon at the 10 nm node and beyond[2-5]. In 2004, graphene, a fascinating material labeled as a perfect two dimensional (2D) crystal with an electron mobility approaching 200,000 $cm^2/Vs$ at room temperature, was realized and has shown promise as a silicon replacement[6-8]. Furthermore, following research has unveiled other similar materials that exist as layered 2D-materials, including BN, $Bi_2Te_3$, $Bi_2Se_3$, $MoS_2$, and etc. These other materials can also be isolated to a single atomic layer via mechanical exfoliation [6, 9-11]. However, researchers have noticed that the deposition of high-k dielectric onto 2D crystals, such as graphene, is not as easy as deposition onto Ge or III-V bulk materials. A typical example is the failure of $Al_2O_3$ deposition on graphene basal plane with trimethylaluminum (TMA) and water as ALD precursors, which is the most reliable ALD process with a wide process window. This failure has been understood to be caused by the difficulty of forming chemical bonds on the graphene basal plane due to existing global $sp^2$-hybridation[12,13]. Despite several successful attempts to integrate high-k dielectrics onto 2D systems[11,14,15], the integration of high-k dielectric onto such 2D crystals has not been thoroughly studied. In this letter, we focus on the growth of ALD $Al_2O_3$ on two typical 2D materials: boron nitride, a sister material of graphene and previously used as a graphene dielectric[16]; and $MoS_2$, a promising layer-structured semiconducting material with a satisfying band gap. Our results show that the initial ALD growth on such 2D materials is determined by physical adsorption of the precursors, and therefore is very sensitive to growth temperature. As a result, the ALD process window for 2D crystals would be consequently reduced. By using the Lennard-Jones potential model, we propose several ways to optimize ALD growth on these 2D crystals, which are different from the methods used on 3D bulk crystals.

BN and $MoS_2$ 2D crystals were thinned from bulk crystals by mechanical exfoliation[6], and then transferred to 300-nm $SiO_2$ covered Si substrates. After being cleaned in solvents to remove tape residue, the samples were loaded into an ASM F-120 ALD system. TMA and water were used as precursors. Pulse times of 0.8 and

1.2 seconds were used for TMA and water, respectively, with a purge time of 6 seconds for both. $Al_2O_3$ was deposited with a range of substrate temperatures from 200°C to 400°C by 50°C steps. Figure 1(a)-1(f) show selected AFM images on BN and $MoS_2$ surfaces after 111 ALD cycles at 200°C, 300°C and 400°C, with an expected $Al_2O_3$ thickness of ~10 nm. The $Al_2O_3$ growth rate on $SiO_2$ substrates did not have significant temperature dependence; however, its growth on BN and $MoS_2$ flakes was strongly temperature dependent. We observed a uniform $Al_2O_3$ layer formed at 200°C on both BN and $MoS_2$ substrates. Our previous study showed that the leakage current density was relatively small (~$2\times10^{-4}$ A/cm$^2$ under 1V gate bias) for $MoS_2$ based metal-oxide-semiconductor structure, suggesting that the ALD $Al_2O_3$ thin film on $MoS_2$ was of good quality[17]. With elevated growth temperatures, it was obvious that the $Al_2O_3$ film was not uniform on both BN and $MoS_2$ substrates. When the growth temperature was increased to 250°C, pinhole defects started to appear at the 2D surface. With further increase of growth temperatures, these pinholes tended to expand and finally connect with each other, leaving island like $Al_2O_3$ clusters on the 2D basal plane. In contrast to the growth on basal plane, the growth on edges remain constant at the range between 200℃ to 400℃, due to the existence of dangling bonds at the basal edges[12, 13]. We analyzed AFM data with a MATLAB script to quantify the $Al_2O_3$ coverage, and used this as a metric for the ease of ALD growth, although the coverage percentage may have evident run-to-run variance due to fluctuations of chamber pressure, which has a significant impact to the surface adsorption. The Otsu method was applied to distinguish the boundary between the regions on BN or $MoS_2$ flakes "with" or "without" $Al_2O_3$ growth[18]. As shown in Figure 2, the $Al_2O_3$ coverage was monotonically decreasing with increased temperature, and had a slightly increased coverage ratio on $MoS_2$ than that on BN at higher temperatures. Such temperature dependent growth indicates that the growth is controlled by physical adsorption of the precursors at the substrate surface, and will be further discussed later.

In order to achieve insight in the understanding of the interactions at the substrate surfaces and hence understand the initial ALD cycles for 2D crystals, density function theory (DFT) studies were performed by using the M06-2x method[19] with basis sets 3-21G(d) for Mo and 6-311G+(d,p) for H, O, C, B, N, and Al. Table 1 shows the calculated adsorption energies of the two ALD processes. It can be seen that the binding energy of TMA on BN is 8.7 kcal/mol greater than that of

$H_2O$ on BN and the binding energy of TMA on $MoS_2$ is 21.9 kcal/mol greater than that of $H_2O$ on $MoS_2$. This implies that TMA is more easily physically absorbed on both types of crystals. Figure 3 shows the four adsorption structural models. For the BN system, the O atom of $H_2O$ is adsorbed at the B atom while the Al atom of TMA is adsorbed at the N atom. The calculated distances of O-B and Al-N are 2.769 and 2.604 Å, respectively. These distances are greater than the corresponding covalence bonds. Therefore, the adsorption energies include the front molecular orbital interaction and van der Waals contribution. For $MoS_2$, $H_2O$ and TMA are all adsorbed at the S atom. The length of Al-S bonds is predicted to be 2.653 Å while the shortest length of O-S bonds is 2.833 Å. Apparently the distances of Al-S and Al-N are shorter than that of O-S and O-B, respectively, though the atomic volume of Al is greater than that of O. Therefore, compared to the $H_2O$ adsorption, the TMA adsorption is more stable.

Table 2 lists the atomic charges, polarizabilities, and frontier molecular orbital levels of the interaction atoms in $H_2O$, TMA, BN, and $MoS_2$. It can be seen that O atoms with negative charges would have electrostatistic interactions with the positively charged B and Mo atoms, while Al atoms with positive charges would be interacting with the negatively charged N and S atoms. Also, one can see that the polarizability of TMA is much greater than that of $H_2O$, while the polarizability of $MoS_2$ is much greater than that of BN. This implies that the interactions of TMA-$MoS_2$ would have the largest dispersion energy and the interactions of $H_2O$-BN would have the least. In addition to this van der Waals interaction, the frontier molecular orbitals of these model molecules may take an important role in the combination. From Table 2, we see that the gaps between the LUMO and HOMO level are 0.5942 au for $H_2O$, 0.3142 au for TMA, 0.3335 au for BN, and 0.1151 au for $MoS_2$. Thus, the orbital interactions of $H_2O$ with BN and $MoS_2$ would be less than that of TMA with BN and $MoS_2$, respectively. This analysis supports the predicted result of adsorption energies.

In classical ALD theories, deposition with precise thickness control is determined by self-limited precursor adsorption at substrate surfaces, and is classified into two types: physical and chemical adsorption surface. During the initial ALD cycles on 2D crystals, excepting a few chemically active materials such as the topological insulators $Bi_2Te_3$ and $Bi_2Se_3$ which are easily oxidized at growth temperatures and hence facilitate the formation of chemical bonds for precursors at the surface[14], chemical adsorption is rarely observed. This is due to the absence of dangling bonds at their basal planes. Consequently, physical adsorption is the dominant adsorption method at the 2D surface. This view is also supported by the result that such deposition is strongly temperature dependent. It is interesting to note that ALD $Al_2O_3$ can be deposited on BN at 200°C, while $Al_2O_3$ can only grow at graphene's edges, even though BN is extremely structurally similar to graphene. Such a difference between $Al_2O_3$ deposition on graphene and BN can be explained using the framework of the Lennard-Jones potential model, which has been generally used to model the molecular adsorption on graphene and carbon nanotube surfaces[20-22]. As shown in Figure 4, for each ALD pulse-purge cycle, the pulse action pushes the precursor molecules to the vicinity of substrate where the molecule has the lowest potential energy; while the purge action pushes the molecule away from the substrate, to the x-axis infinity, where the molecules encounter an energy barrier. There are two factors that determine the ultimate molecular state: One is the depth of the potential well, shown as the adsorption energy and determined by the polarizability of the substrate and molecules. Using BN as an example, nitrogen serves as a positive charge center while boron serves as a negative charge center, while graphene has no polarization due to perfect symmetry, the interaction between the BN molecule and ALD precursors would be stronger than that of graphene and ALD precursors. That is to say, the depth of the potential well in the BN system will be larger than that in the graphene counterpart. The other reason is the growth temperature, which is near the thermal energy of the precursor molecules. At lower temperatures, the thermal energy is small enough that the molecules are trapped in the potential well, despite the purge action, while at higher temperatures where the thermal energy of the molecule is greater than the depth of the potential well, the excited molecule can escape, thus undoing ALD cycles.

Therefore, the ALD window for deposition on 2D crystals is different from previous studies on bulk materials. For bulk substrates, the lower temperature limit of the ALD growth window is determined by precursor condensation and incomplete reaction at lower temperatures, and the high temperature limit is determined by precursor decomposition as well as desorption[23]. For 2D substrates, the low temperature limit still remains similar as it's only related to the precursors, regardless of the substrate material. However, the high temperature limit, since desorption is much easier at 2D surfaces, is at a dramatically lower temperature. This creates a large challenge to dielectric integration for high performance devices, such as threshold shifts observed in our previous study on $MoS_2$ top-gated MOSFETs[17].

Given our discussion above, we can clearly see that the first several ALD cycles is critical, not only for properties related to interface quality, but to allowing further deposition as $Al_2O_3$ can provide dangling bonds for the chemical adsorption of the precursors. One way to optimize the ALD process is to change the pulse and purge times in to better control the surface adsorption/desorption at the initial stages of deposition. Alternatively, a seeding layer, such as an ultrathin Al film, or $Al_2O_3$ film deposited by ALD at lower temperature, can also provide a solution for high quality dielectric growth as played with graphene[24]. Another related question is the "interface" issue, as it has been generally accepted that the origin of interface states is attributed to unpassivated dangling bonds[25]. In the absence of dangling bonds at surface of 2D crystals, the definition of "interface states" between 2D crystals and dielectrics may need to be reconsidered, and such an issue may need further investigation. This research is still on-going.

In summary, we have demonstrated direct ALD growth of $Al_2O_3$ with TMA and water as precursors on 2D crystals of BN and $MoS_2$. We have also performed a DFT study on surface adsorption of 2D crystals at different geometric substrate locations. Both experimental and theoretical results show that the ALD growth on 2D crystals is determined by physical absorption, and is enhanced by in-plane polarization of the substrate. Our results have shown provided insight into growth mechanisms and will allow better solutions for high-quality dielectric integration to be found, and provides a big step forward for novel devices based on 2D crystals in the future.

The authors would like to thank G. Q. Xu, H. B. Lu, N. J. Conrad and P. Kim for valuable discussions.

Figure Captions:

Figure 1: AFM images of BN or $MoS_2$ surface after 111 cycles of ALD $Al_2O_3$ at 200℃, 300℃ and 400℃. All images are taken in a 2 μm by 2 μm region with a scale bar of 500 nm.

Figure 2: $Al_2O_3$ coverage estimation from MATLAB analysis by Otsu method.

Figure 3: Binding structure models of $H_2O$ and TMA on BN and $MoS_2$. Bond length is in Å.

Figure 4: An illustrative Lennard-Jones potential Model for physical adsorption at 2D crystal surfaces. Strong, intermediate and weak adsorptions are qualitatively presented here. The depth of the potential well indicates the adsorption energy, noted as $E_{ads}$. These curves are not scaled by calculated values.

Table 1.   Binding energy ($E_{ads} = E_a + E_b - E_{ab}$) in kcal/mol

| adsorption | $E_{ads}$ |
|---|---|
| $BN+H_2O \rightarrow BN-H_2O$ | 4.1 |
| $BN+TMA \rightarrow BN-TMA$ | 12.8 |
| $MoS_2+H_2O \rightarrow MoS_2-H_2O$ | 11.1 |
| $MoS_2+TMA \rightarrow MoS_2-TMA$ | 33.0 |

Table 2. Atomic charges, polarizabilities, and frontier molecular orbital levels of the interaction atoms in $H_2O$, TMA, BN, and $MoS_2$ in atomic unit.

| | $H_2O$ | | TMA | | BN | | $MoS_2$ | |
|---|---|---|---|---|---|---|---|---|
| $\varepsilon_i$(LUMO) | 0.1933 | | -0.0068 | | 0.0363 | | -0.1672 | |
| $\varepsilon_i$(HOMO) | -0.4009 | | -0.3210 | | -0.2972 | | -0.2823 | |
| $P_i$ | 6.140 | | 48.996 | | 34.666 | | 112.316 | |
| | O | H | Al | C | B | N | Mo | S |
| $Q_\alpha$ | -0.582 | 0.291 | 1.337 | -0.481 | 0.991 | -0.924 | 2.176 | -1.088 |

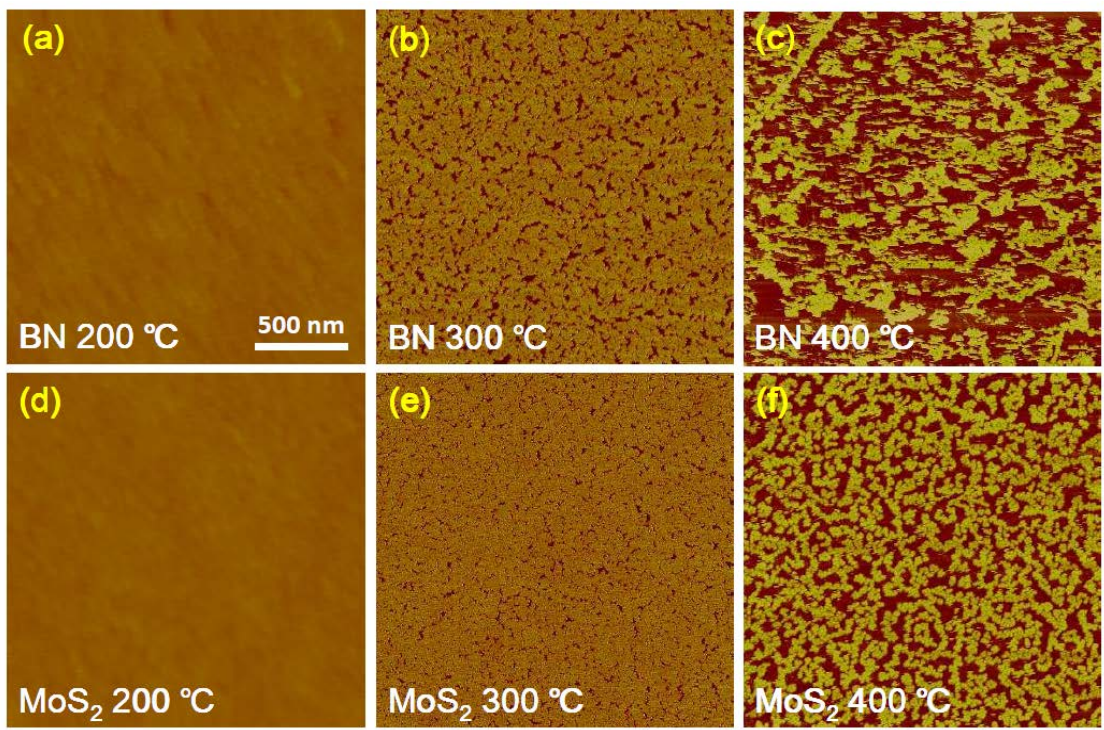

Figure 1

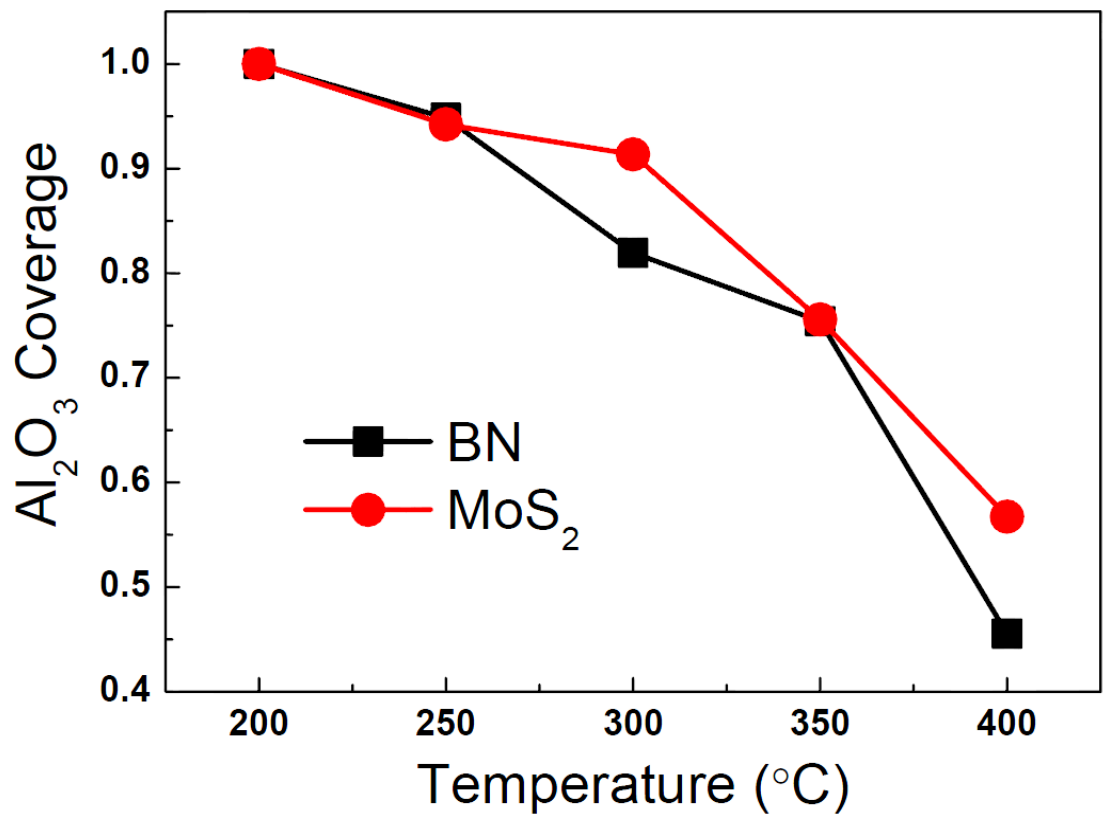

Figure 2

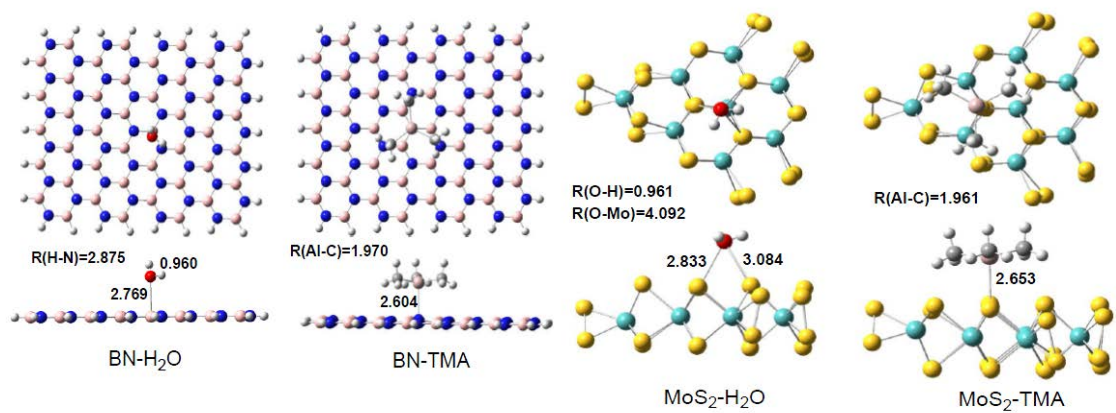

Figure 3

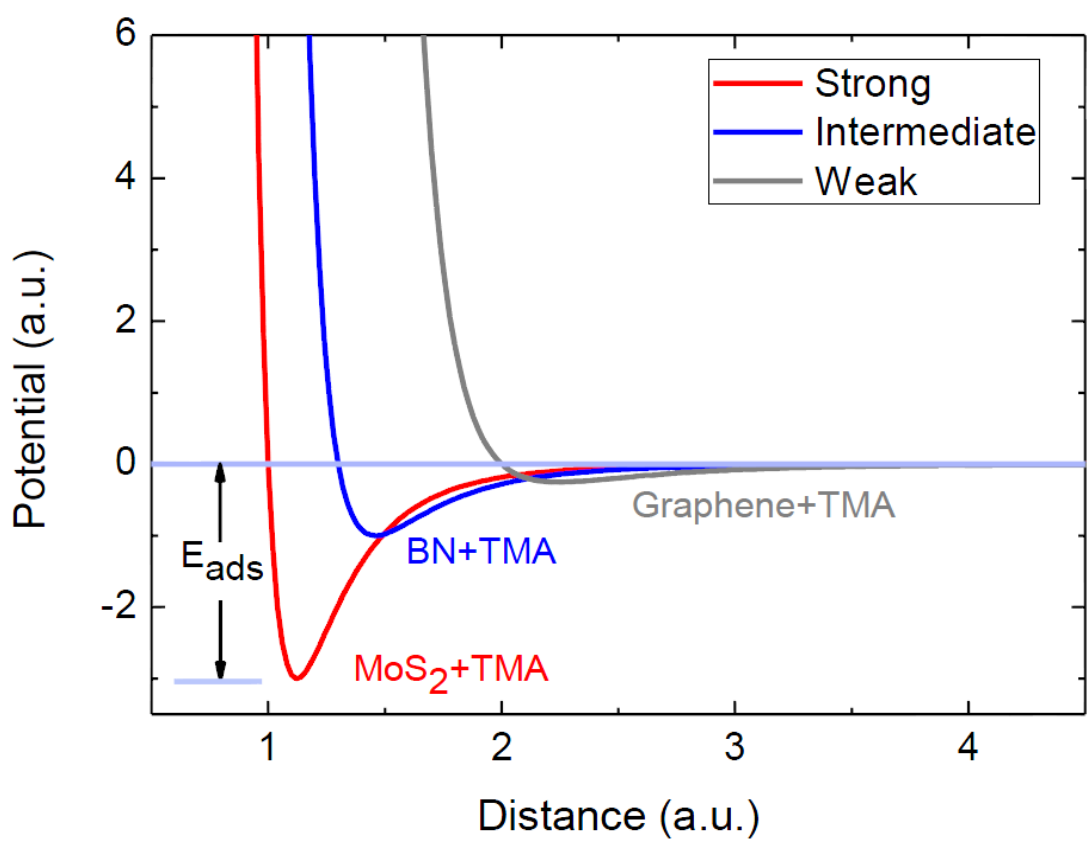

Figure 4